\documentclass[useAMS,usenatbib]{mn2e}
\usepackage{epsfig}
\usepackage{natbib}
\usepackage{subfigure}
\usepackage{lscape}
\usepackage{array}
\usepackage{amssymb,amsmath}
\DeclareMathVersion{bold}

\title[Millisecond Radio Transients]{The Millisecond Radio Sky: Transients from a Blind Single Pulse Search}

\author[Burke-Spolaor and Bailes]{ 
S. Burke-Spolaor,$^{1,2}$\thanks{Email: sburke@astro.swin.edu.au}
M. Bailes$^1$\\
$^{1}$Swinburne University of Technology Centre for Astrophysics and Supercomputing, Hawthorn VIC, Australia\\
$^{2}$Australia Telescope National Facility, CSIRO, P.O. Box 76, Epping, NSW 1710, Australia\\
}
\date{}
\begin{document}

\maketitle
\begin{abstract}
We present the results of a search for transient radio bursts of between 0.125 and 32 millisecond duration in two archival pulsar surveys of intermediate galactic latitudes with the Parkes multibeam receiver. Fourteen new neutron stars have been discovered, seven of which belong to the recently identified ``rotating radio transients'' (RRATs) class. Here we describe our search methodology, and discuss the new detections in terms of how the RRAT population relates to the general population of pulsars. 
The new detections indicate (1) that the galactic z-distribution of RRATs in the surveys closely resembles the distribution of pulsars, with objects up to 0.86~kpc from the galactic plane; (2) where measurable, the RRAT pulse widths are similar to that of individual pulses from pulsars of similar period, implying a similar beaming fraction; and (3) our new detections span a variety of nulling fractions, and thus we postulate that the RRATs may simply be nulling
pulsars that are only ``on" for less than a pulse period. Finally, the newly discovered object PSR J0941--39
may represent a link between pulsars and RRATs. This bizarre object was discovered as an RRAT, but in
follow-up observations often appeared as a bright ($\sim$10 mJy) pulsar with a low nulling fraction. It is
obvious therefore that a neutron star can oscillate between being an RRAT and a pulsar. Crucially, the
sites of the RRAT pulses are coincident with the pulsar's emission, implying that the two emission mechanisms
are linked, and that RRATs are not just pulsars observed from different orientations.

\end{abstract}
\section{Introduction}\label{sec:intro}
The transient radio sky has revealed many classes of exotic objects since the
discovery of pulsars in late 1967 \citep{hewishetal68}. While the first
$\sim$50 pulsars were discovered by searching for their single-pulse emission,
the advent of filterbanks and digital signal processing 
searching in the sub-second radio sky to use more efficient and successful
allowed the use of Fourier search techniques, which are more sensitive to
periodic objects than single pulse searches and have since permitted the
discovery of more than 1500 pulsars. However, the recent discovery of sparsely
bursting rotating radio transients (RRATs, \citealt{mclaughlinetal06}), and a
one-off burst tentatively identified as having an extragalactic origin
\citep[][]{ERB} has spurred renewed interest in the search for single-pulse
emission after 30 years of Fourier transform searches dominating surveys for
pulsars.

The more recent history of single-burst searches commenced with the search of
\citet{shaun} for impulsive signals between 0.001 and 800~ms using a transient
event monitoring system at 843~MHz at the Molonglo cylindrical telescope.
While they found only pulsars and ambiguous events, they were able to put an
upper limit on the event rate of $1-25$~ms duration events of
0.017~events\,\,s$^{-1}$\,sr$^{-1}$ for pulses with energy density $>
10^{-28}$~J\,m$^{-2}$\,Hz$^{-1}$. \citet{nice99} also recognised that some
pulsars might be undetectable in periodicity searches and subsequently
re-searched data from a 68 deg$^2$ area of sky along the galactic plane for
single pulses with duration between $0.5-8$~ms, revealing one new, slow
(2.3~s) pulsar. The survey which discovered RRATs \citep[][hereafter
M06]{mclaughlinetal06} was the first large-scale examination of a significant
data set for impulsive bursts since the early pulsar surveys, while the search
of the Magellanic Cloud surveys \citep{ERB} and a transient search with
Arecibo telescope \citep[][hereafter D09]{palfa} and a deeper search of the
M06 data \citep[][hereafter K09]{evanrrats} followed using similar techniques.

There are several known sources of fast transient radio emission which we
expect to detect as single pulses on sub-second time scales. Periodic rotators
which have irregular emission properties include the aforementioned RRATs,
which are a form of rotating neutron star, emitting pulses very sparsely at
several to many times the rotation period of the star. At present it is
unclear from the few observed RRATs how they relate to the neutron star
population as a whole, or how they may relate to other sub-populations of
neutron stars such as anomalous X-ray pulsars or magnetars.
It is possible that they represent only a phase through which all pulsars
undergo when they reach a sufficient age \citep[e.g.][]{mauranewpaper,evan},
or that are simply distant, low-luminosity pulsars with extreme pulse
amplitude modulation \citep{patrick06}. Such distinctions are important for
reconciling the so-called ``birthrate problem''; it has been recently noted
that the net neutron star birthrate calculated from apparently disparate
neutron star populations (X-ray dim isolated neutron stars (XDINS), magnetars,
RRATs, pulsars) exceeds the rate of the core-collapse supernovae that are
believed to produce them \citep{popov06,evan}. Recent works are endeavouring
to perform analyses that make the distinction clearer from pulse timing
(e.g. \citealt{mauranewpaper}).

High pulse-to-pulse modulation and periodic nulling represent types of signal variance observed in the single pulse emission of some pulsars \citep[e.g.][]{herfindalrankin07,sometimesapulsar}. Fourier transform searches become less sensitive to such sources due to the objects' erratic emission and the often short duty cycles and long rotation periods of the stars. Such neutron star-related phenomena promise insights into the localisation and power source of radio pulsar emission. Single pulse searches have the ability to recognise these sources, and may serve to reveal a bridging population or define a distinction between regular pulsars that emit steadily and the seemingly more rare ``part-time'' pulsars that only emit detectable emission some fraction of the time, revealing whether RRATs are (in physical mechanism) just extreme examples of the nulling and modulation seen in some pulsars. 

Theoretical arguments predict fast transients from a variety of non-neutron star origins, such as coalescing systems of relativistic objects \citep{hansenlyutikov01}, evaporating black holes \citep{rees77}, or supernova events \citep[e.g.][]{colgatenoerdlinger71}. The aforementioned detection of a single, apparently extragalactic, 5~ms and 30~Jy burst detected by \citet{ERB} indicated that the detection of such sources in other galaxies might be possible. While the Lorimer et al. detection was the primary motivator for performing our search, fourteen other discoveries associated with intermittent neutron star phenomena resulted, and these are presented and analysed below. No further detections were made of bursts attributable to an extragalactic origin. Fifteen seemingly non-astrophysical, frequency-swept bursts were discovered with a similar frequency-dependent sweep rate (or dispersion measure) to the Lorimer et al. detection (except that they occurred in all 13 instead of just 3 beams of the Parkes multibeam receiver), however a deeper analysis and the interpretation of those signals will be presented elsewhere (see Burke-Spolaor et al., in prep). 


This paper presents the detections from a blind search for single pulses in more than 900 hours of archival data from the pulsar surveys of \citet{ED} and \citet{BJ}. The data is described in section \ref{sec:sample}, and the search methodology in section \ref{sec:methods}. The fourteen new objects and other results uncovered by this search are presented in section \ref{sec:results}. The implications of these detections are discussed in section \ref{sec:discussion}.

\begin{figure*}
{
 \includegraphics[trim=0mm 5mm 15mm 0mm,clip,width=0.75\textwidth]{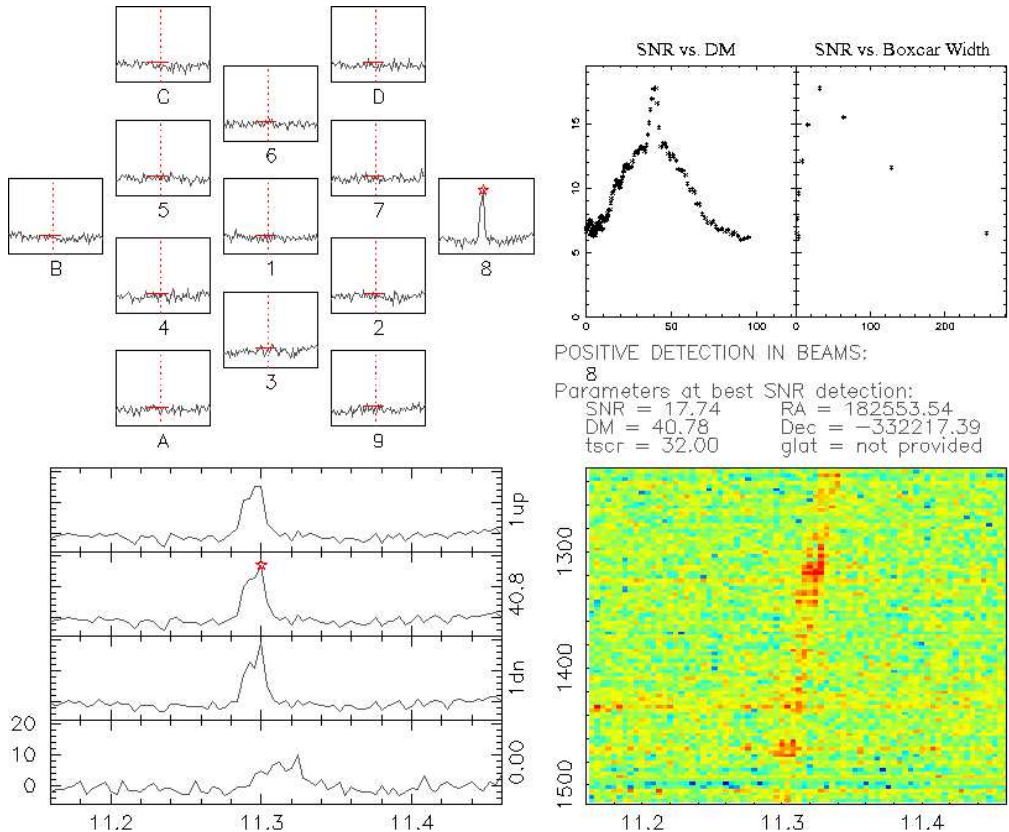}
}
\caption{A detection plot for a burst from J1825--33. {\sc Upper left:} Time
series from each of the 13 beams dedispersed at the DM of brightest detection;
each field plots the power versus time, normalised by the RMS of the time
series. All fields plot the same abscissa and ordinate range, and the beam
(and sample in the time series) of brightest detection is indicated by a red
star. In the other beams, the sample of peak detection is marked by a vertical
dotted line, and the detected width of the pulse is marked by a horizontal
bar. {\sc Lower left:} Time series for the maximum-SNR beam dedispersed at the
DM of brightest detection, stacked with the time series from one DM trial
higher (``1up''), one trial lower (``1dn''), and the DM~$=0$ time series. As
in the upper left panel, the ordinate gives normalised power (SNR), while the
abscissa gives time since the beginning of the file (in seconds). {\sc Bottom
right:} A false colour image of the received power as a function of frequency
(in MHz) versus time since the beginning of the file (in seconds), revealing
the frequency-dependent structure of the detected pulse in the beam of
brightest detection. {\sc Upper right:} Signal-to-noise ratio as a function of
DM trial and boxcar filter size, and best-fit information about the detection
(DM, SNR, pointing position of the beam of brightest detection, and boxcar
width ``tscr''). Pulses with resolved complex structure can show multiple
peaks and other structure in their SNR vs. DM curve, though genuinely
dispersed pulses (and the curve above at the correct DM$\pm5$) generally obey
the function given by \citet{cordesmclaughlin}.}
\label{fig:viewsus}
\end{figure*}

\section{Data sample}\label{sec:sample}
The data used in this search were the southern-sky ($-100^\circ<l<50^\circ$) intermediate \mbox{($5^\circ<|b|<15^\circ$)} and high \mbox{($15^\circ<|b|<30^\circ$)} galactic latitude surveys of \citet{ED} and \citet{BJ}, respectively;  here $b$ and $l$ represent galactic latitude and longitude, respectively. 
Both surveys employed the same analogue 1-bit filterbank as a back-end to the 20cm multibeam receiver system on the Parkes Radio Telescope. The multibeam receiving system allows the simultaneous observation of 13 positions on the sky with approximately 30 arcminutes between beam positions \citep{multibeam}. The twelve outer beams are arranged in two concentric sets of six (see the upper left panel of \mbox{Fig. \ref{fig:viewsus})}, with a full-width half power beamwidth of each beam at 14 arcminutes. Due to the sharp sensitivity fall-off within 24 arcminutes of each beam's pointing centre, a pointlike astrophysical signal will generally be detected in no more than four beams of the receiver, and unless particularly bright, usually just one or two.

Each survey sampled 96 channels across a 288~MHz band centred on a frequency of 1372.5~MHz. The data were recorded after the polarisations were summed and then sampled at 125~$\mu$s. Each 13-beam pointing of the telescope lasted 4.4 minutes. The data total 12,090 beam hours (there are 13 beams), taken over the years 1998-1999 (Edwards) and 2001-2002 (Jacoby). The total number of pulsars detected in the periodicity searches of these surveys were 170 for Edwards et al. and 62 for Jacoby et al, giving a total of 230 independently detected pulsars.

All follow-up data presented here were taken with the new 2-bit digital filterbank system currently installed at Parkes Radio Telescope that use CASPER tools and is known as BPSR. Data with this system were taken with 1024 channels across 400~MHz centred at 1381.8~MHz, and sampled at 64~$\mu$s. Filters blocking emission from a geosynchronous satellite visible by the telescope render frequencies above 1520~MHz unusable.

\section{Single burst search process}\label{sec:methods}
The most significant difficulty in single pulse searches is allaying the millions of spurious candidates which arise from local man-made and natural radio frequency interference, such as lightning. Our search employs various stages of automatic radio frequency interference (RFI) elimination before candidates are manually assessed. We generally followed the steps of \citet{cordesmclaughlin}, however with adjustments in the techniques used for RFI rejection, matched filtering, and visual inspection. The main discriminant used in single pulse searches is the fact that astrophysical transients exhibit a well-defined frequency-dependent delay caused by the propagation through multiple ionised plasma screens in the interstellar medium, visible as a time lag across the bandwidth of our instrument.
The dispersion measure, DM, measures the integrated line-of-sight column density of free electrons in units of pc\,cm$^{-3}$. Locally generated signals generally do not appear dispersed (DM~$\sim 0$), while genuine bursts have a decrease in measured significance as the trial dispersion increases from the true dispersion of the source. Periodic sources will show multiple detections at the same dispersion measure. The radio interference ``forest'' local to any given astronomical observing site gives rise to the need for a set of thresholds that remove broad, zero-dispersion signals. We therefore have increased confidence in candidates at higher dispersion.

Our detection thresholds were estimated by manual inspection of the data and trial runs on a sub-sample of data to characterise the typical local RFI environment of the Parkes Telescope. These tests indicated that events separated by at least 3.75 ms (30 data bins at the native time resolution) could be considered independent, that signals peaking in signal-to-noise ratio (SNR) below DM~$=5$ were most likely RFI or statistically insignficant, and that setting a search threshold at six times the RMS noise of the dedispersed time series was sufficient to filter out the majority of candidates which were due to statistical fluctuations or narrow band interference.

The search process followed several steps. At each trial DM, after
dedispersion the frequency channels were summed and a search for SNR$\geq 6$
pulses was performed of the complete 4.4 minute observation. An increased
sensitivity search for broader pulses was then performed, using a progressive
box-car smoothing function ranging $2^1, 2^2,...,2^8$ time samples,
corresponding to a maximum filter width of 32~ms. The 32\,ms maximum width was
chosen to significantly decrease the ratio of local to astrophysical
candidates (as determined by tests on a data subset), and to preserve our
sensitivity to impulses of low dispersion measure (the total delay in
milliseconds across our band is roughly equivolent in value to the dispersion
measure of an impulse in units of pc\,cm$^{-3}$). We have decreased
sensitivity to pulses of half-maximum width, $w$, outside the range
$0.125<w<32$~ms. Additionally, the one-bit nature of the data made it
difficult to discover pulses with width less than a sample period, as we only
had $\leq 96$ functioning one-bit channels, and an RMS that meant it was
difficult to have much confidence before summing samples. We searched in DM
from $0<{\rm DM}<600$~pc\,cm$^{-3}$, which for the galactic latitudes of the
survey is sufficient to probe out to at least 50~kpc except in the regions a
few tenths of a degree from $l=0, |b|=5$.

For each beam, a first-pass RFI rejection was performed by doing temporal
coincidence matching for all trial dispersions and matched filter steps; in
this way we were able to reconstruct multi-plane information about
singly-detected events.  A further temporal coincidence search of all
remaining events was done in the 13 beams, conservatively rejecting all
candidates which had detections coincident in more than nine. After this step,
we have a full event archive in which are encoded the widest duration
statistics for detected beams, and information about the beam, DM, and
smoothing filter of strongest detection. Event archives which contained less
than three events, and those which had a peak in signal-to-noise at less than
DM~$=5$, were rejected.  This step significantly reduced the number of high
dispersion candidates which were due purely to narrow-band RFI and small
number statistics. The information in the remaining event archives were used
to make the plots used in manual candidate assessment.


\begin{figure*}
{
 \includegraphics[trim=3mm 20.5mm 3mm 25mm,clip,width=0.8\textwidth]{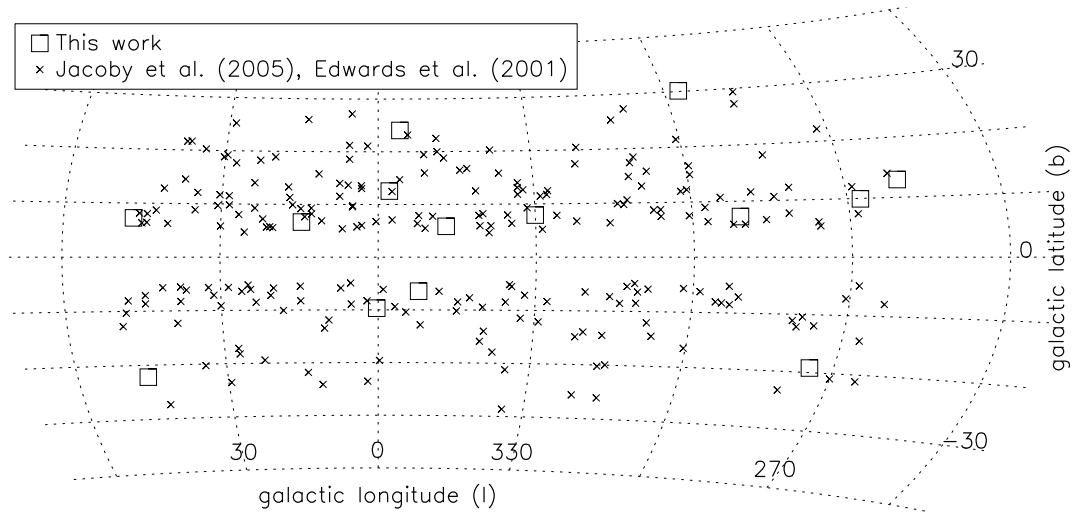}
}
\caption{A galactic map of the pulsars originally detected in the data, plotted with the new detections presented in this work.}
\label{fig:galmap}
\end{figure*}

\begin{table*}
  \centering
  {\small
  \begin{tabular}{lrrccccccccc}
\hline
\textbf{J2000}&\multicolumn{1}{c}{\textbf{l}}&\multicolumn{1}{c}{\textbf{b}}& \textbf{DM}  &${\bf d}$ &${\bf N}_{\bf p}/{\bf t}_{\rm {\bf obs}}$& \textbf{P}& $\bf{w}_{\rm \bf{i}}$ &  $\bf{w}_{\rm \bf{s}}$  &$\bf{S}_{\rm \textbf{max}}$ \\
\textbf{Name}&(deg) & (deg)&(pc\,cm$^{-3}$)&  (kpc)     &(hr$^{-1}$)                                                & (ms)        &(ms) & (ms) & (mJy)\\
 \hline
 \hline
J0735--62$^\dag$& 274.7 &-19.2& 19.4$\pm$7.8 & 0.93 &--- & 4862.940(2)          & 37.9 & 14 & 160  \\ %
J0923--31$^\#$     & 259.7 & 13.0& 72$\pm$20     & 0.55 &1.7& ---                              & 30(213)& 30(213) & 120 \\ 
J0941--39$^\dag$*& 267.8 & 9.9  & 78.2$\pm$2.7 & 0.56 &--- & 586.77841838(3)& 105.6& 6 & 580 \\ 
J1129--53               & 290.8 & 7.4  & 77.0$\pm$2.5 & 2.05 &36.2&   1062.880(5)     & 19.1 & 14 & 320 \\ 
J1226--32               & 296.9 & 30.1& 36.7$\pm$9.9 & 1.42 &40.9*& 6193.1(5)            & 60.7 & 26  & 270\\ 
J1534--46$^\dag$ & 330.0 & 7.9  & 64.4$\pm$7.8 & 1.75 &--- &  364.835(9)           & 25.5 & 10  & 140\\ %
J1610--17$^\#$      & 355.6 & 24.1& 52.5$\pm$3.0& 2.10 &13.6*& ---                          & 5       & 5    & 230\\ %
J1647--36                & 347.1 & 5.8  &223.9$\pm$1.6&5.23 &436.4*& 210.726(1)     & 9.9    &9     & 250\\ 
J1654--23                & 357.9 & 12.6& 74.5$\pm$2.5& 2.04 &40.9*& 3986.48(5)        & 0.52 &0.52& 710\\ 
J1753--38                & 352.3 & -6.5 &168.4$\pm$1.3& 3.79 &26.3& 621.000(5)         & 4.8   & 2.5 & 440\\ 
J1753--12                & 14.6   & 6.7   & 73.2$\pm$5.2& 1.85 &40.9*& 405.454(1)       & 18.8  & 14 &  160\\ 
J1825--33                & 0.3      & -9.7 & 43.2$\pm$2.0& 1.17 &14.4& 1271.2(2)              & 16.5  & 16 &  360\\ 
J1850+15$^\dag$& 46.7   & 7.3   &24.7$\pm$8.7& 1.79 &---&  1383.965(3)           & 31.8 & 6    &  200\\ %
J2033+00$^\dag$& 46.0   & -22.2& 45.2$\pm$5.9& 2.28 &---& 5012.918(2)          & 95.2  & 38 &  140\\ %
  \end{tabular}
}
\caption{Data for the 14 new sources detected by this search. Columns: (1)
Name based on J2000 coordinates ($\#$ signifies only one burst detected, while
$\dag$ denotes an object that was Fourier-detectable in the original pointings
or in follow-up data; * see \S\ref{sec:0941} for special remarks on
J0941--39); (2,3) galactic longitude ($l$) and latitude ($b$); these
coordinates have an error of approximately $\pm$15 arcminutes due to the poor
positional localization of the multibeam receiver; (4) best-fit dispersion
measure and error; (5) distance to the object based on the DM and NE2001
galaxy electron density model \citep{ne2001}; (6) the pulsation rate based on
all data taken on object position to date, where the pulsar was not detected
above a SNR of 5 in the folded data. An asterisk indicates an object for which
only the 4.4-minute detection pointings have been used to compute this number;
(7) the best fit period to all detected pulses with error on the last digit in
parentheses; (8) integrated pulse width measurement at 50\% of maximum after
the stacking of all detected single pulses. For PSR J0923--31, we quote the
narrow component width and the total pulse width; (9) pulse width measurement
at 50\% of the peak of the brightest detected pulse. PSR J0923--31 specified
as in column 8; (10) peak flux density of brightest detected single
pulse.}\label{table:newstuff}
\end{table*}

The manual candidate ranking involved the viewing of each candidate using the tool displayed in Figure \ref{fig:viewsus}, which shows a single pulse detection from one of our new discoveries, and contains the information with which one can quickly discriminate a true burst from a false detection. During candidate viewing, we simultaneously queried the ATNF pulsar catalogue for pulsars within two degrees of the brightest beam, such that we were able to easily determine DM and position incidence, and attribute detections to previously known pulsars. Non-RFI detections not associated with known pulsars were considered new transient candidates.

Assuming that detections which show repeated pulsations are part of the neutron star population, we fit a rotation period to each object which has emitted at least three detectable pulses. Pulse periods were derived by first fitting periodicities by eye to the detected pulses. The data were then broken into sub-integrations at the estimated period, and sub-integrations containing on-pulses were extracted and used to do an optimised automatic search in period and DM space using the {\sc pdmp} program distributed with the freely-available {\sc PSRchive} software suite\footnote{Available from the web on http://psrchive.sourceforge.net/}. Where follow-up data was available and the observations were of higher SNR, those data were used to compute the fits.


\section{Results}\label{sec:results}
Our blind search yielded the detection of several thousand single pulses with which we could associate 91 pulsars, that were already previously detected through Fourier transform search techniques by Edwards et al. (2001) and Jacoby (2005), indicating that roughly 40\% of the pulsars detected in their periodicity searches have one or more detectable pulses per 4.4 minute pointing. The pulsars we detected do not appear to have any selective distribution in age or magnetic field strength, have periods ranging 0.089 to 8.5 seconds, and appear to have primarily populated the ``island'' of standard pulsars in the period-period derivative diagram with periods between $\sim$0.1 and 3 seconds.

The new objects detected in this search include two sources which emitted only one (albeit very convincing) pulse over the time we have observed them, and 12 sources from which multiple pulses were observed. All of these new sources are interpreted to be erratically emitting neutron stars.
The best-fit parameterisations for each source are presented in Table \ref{table:newstuff}, and a more detailed description for each is given below. The implications of the new detections are discussed in \S \ref{sec:discussion}.

\subsection{New Transient Sources}
The properties of the pulsed sources presented here give no indication that
they are associated with a non-neutron star origin. Below we describe the
properties unique to each new object, and note the amount of follow-up time
which has been dedicated to each. Where emission was detectable in a Fourier
search, the intermittency ratio \mbox{($r={\rm SNR}_{\rm Single
pulse}/{\rm(SNR)}_{\rm Fourier}$,} as given by D09) is provided.

\subsubsection{PSR J0735--62} 
This source emitted several pulses with SNR~$>7$ and many pulses below a SNR
of 6 in the discovery pointing, and after a period was determined from the
brightest bursts, the source was visible when the data were folded at the
manually measured period with a SNR of just below six. The source appears to
experience severe scintillation of its signal due to its low dispersion
measure, and it is likely that because of this the six-minute tracks on this
source over 93 minutes of follow-up observations showed no detectable emission
in single pulses, however showed a weakly visible signal when the data were
folded at the nominal period and DM. Three-period averaged pulses from this
pulsar are shown in Figure \ref{fig:stacks07}. Despite its steady emission,
the long period and narrow duty cycle prevent it from being detected in
standard Fourier searches of short duration \citep[e.g.][]{palfa}. The
intermittency ratio of this object is $r=1.45$.

\subsubsection{PSR J0923--31}
This source emitted only one pulse in the discovery observation, which
appeared at nine times above the system noise and with complex profile
structure with a bright, narrow main component. This pulse is shown in Figure
\ref{fig:singlepulsers09}, and appears to have a faint, time-resolved, double
peak. Subsequent pointings, totalling 30.9 minutes, revealed no further
emission from this location. Further pulse detections are required to rule out
local interference as the source of the burst, however the appearance in only
one beam of the multibeam receiver and frequency sweep that closely follows
the cold plasma dispersion relation gives a strong implication that the source
is astrophysical.  Based on a 10\% duty cycle and a pulse width of 213~ms, if
this object is a pulsar its estimated period should be $>2$~seconds.

\subsubsection{PSR J0941--39}\label{sec:0941}
This pulsar was discovered after the detection of five single pulses to which
there was no clear periodicity. Follow-up observations of the source
(totalling more than 3 hours spread over approximately one-half year) later
revealed extraordinary behaviour, sometimes appearing with sporadic RRAT-like
emission at a rate of $\sim$90-110 per hour, and at other times emitting as a
bright pulsar with strong pulse-to-pulse modulation and quasi-periodic nulling
(see Fig. \ref{fig:0941}). The pulsar has a triple-peaked profile and broad
duty cycle, with complex sub-pulse drift, mode-changing, and
longitude-dependent modulation when in an ``on'' state (it is these behaviours
which made single-pulse periodicity fitting difficult). A timing program is
underway on this pulsar at Parkes Telescope, and observations of the
properties of this object will be presented in detail in a forthcoming paper.

\begin{figure}
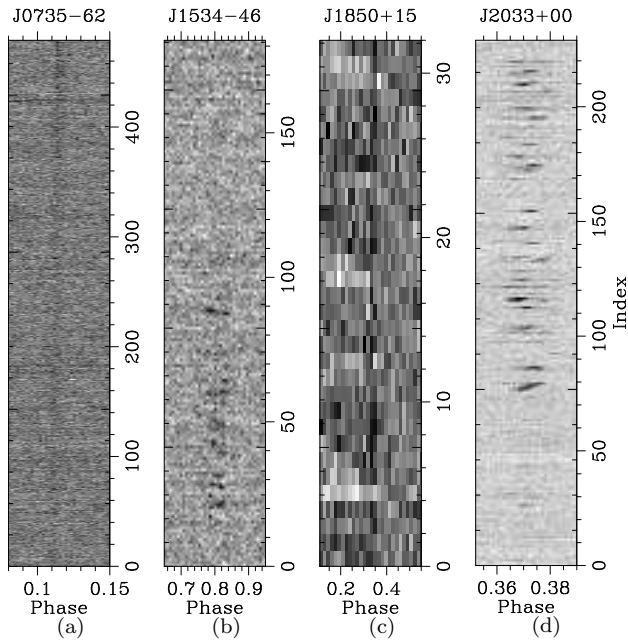

\begin{centering}
 \subfigure[]
{
 \includegraphics[height=0.1\textwidth,trim=0mm 15.3mm 0mm 5mm,clip,angle=270]{Figs/J0735-62stack.ps}\label{fig:stacks07}
}
\subfigure[]
{
 \includegraphics[height=0.1\textwidth,trim=0mm 15.3mm 0mm 5mm,clip,angle=270]{Figs/J1534-46stack.ps}\label{fig:stacks15}
}
\subfigure[]
{
 \includegraphics[height=0.1\textwidth,trim=0mm 15.3mm 0mm 5mm,clip,angle=270]{Figs/J1850+15stack.ps}\label{fig:stacks18}
}
\subfigure[]
{
 \includegraphics[height=0.113\textwidth,trim=0mm 15.3mm 0mm 0.4mm,clip,angle=270]{Figs/J2033+00stack.ps}\label{fig:stacks20}
}

\caption{Greyscale representations of time (or pulse index, y-axis) versus de-dispersed pulse phase---``pulse stacks.'' These four objects were alone in showing Fourier-detectable emission underlying their single bursts. PSR J0735--62 has been folded over three rotation periods, PSR J1850+15 over six periods, and the others show single pulses.}
\label{fig:stacks}
\end{centering}
\end{figure}

\begin{figure}
\begin{centering}
 \subfigure[PSR J0923--31]
{
 \includegraphics[height=0.41\textwidth,trim=0mm 0mm 0mm 0mm,clip,angle=270]{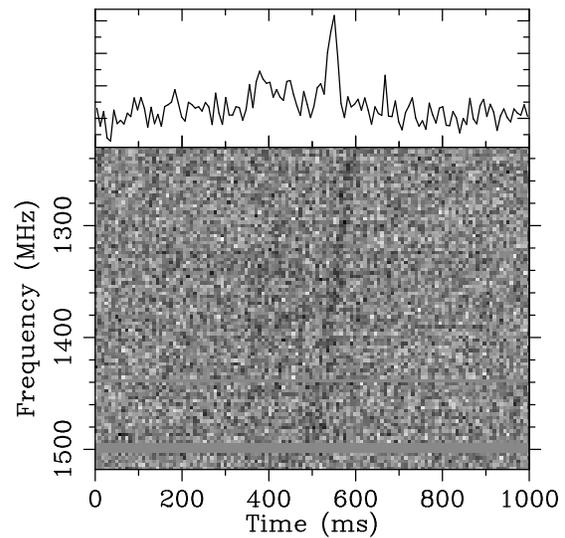}\label{fig:singlepulsers09}
}
 \subfigure[PSR J1610--17]
{
 \includegraphics[height=0.41\textwidth,trim=0mm 0mm 0mm 0mm,clip,angle=270]{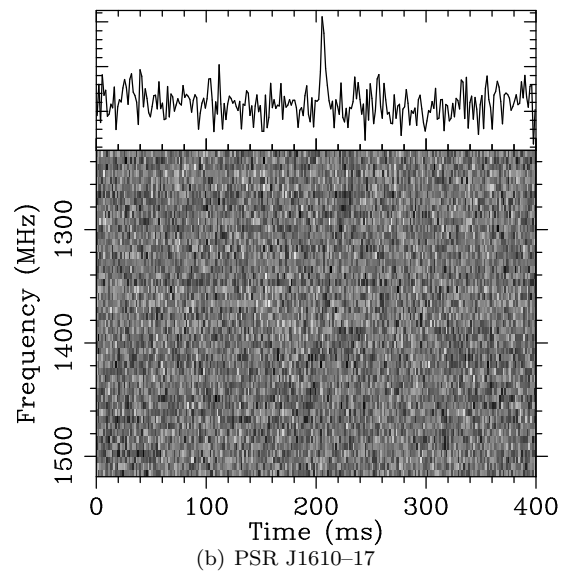}\label{fig:singlepulsers16}
}
\caption{Greyscale spectra (bottom panel) and dedispersed, frequency-integrated timeseries (top panel), showing the single pulses of the two objects from which no further emission has yet been detected.}
\label{fig:singlepulsers}
\end{centering}
\end{figure}

\subsubsection{PSR J1129--53}
This source was originally detected as two single bursts occurring at the same
DM in independent pointings separated by 14 arcminutes. Follow-up observations
totalling 15.5 minutes at a position half-way between the original two
detection positions (it is this position listed in Table \ref{table:newstuff})
revealed 10 further pulses which were extracted and used to fit a period and
DM. The individual pulses from this object exhibit a variety of pulse shapes,
with some pulses exhibiting evenly double-peaked emission.

%

\subsubsection{PSR J1226--32}
Three pulses of SNR~$>12$ were detected from this object during the discovery
pointing, and a best-fit to these pulses gave us the period and dispersion
measure quoted in Table \ref{table:newstuff}. When folded at the fitted
period, the rotations which did not contain pulses of SNR~$>12$ showed no
detectable emission. We have not yet had the benefit of follow-up observations
for this object, however the three dispersed pulses emitted by the object show
consistency in both dispersion sweep and frequency-dependent modulation,
indicative of the bursts' propagation through the interstellar medium from a
common origin.

\begin{figure*}
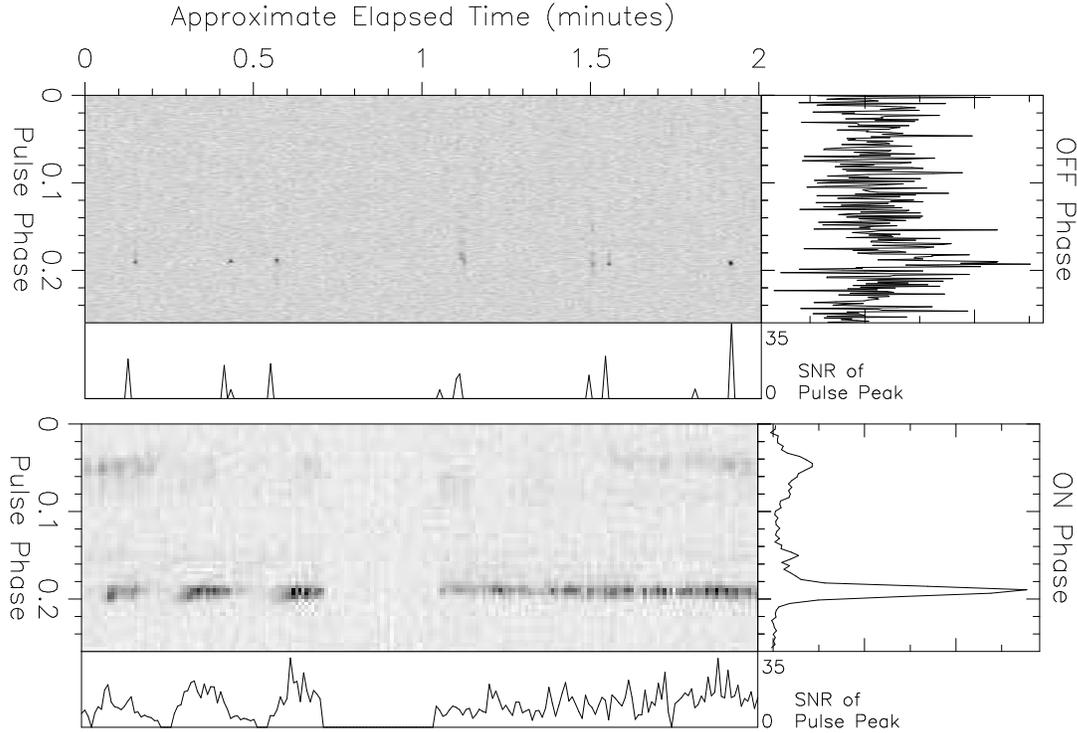

\begin{centering}
 \subfigure
{
 \includegraphics[angle=180,width=0.8\textwidth,trim= 0mm 0mm 0mm 0mm,clip]{Figs/off.ps}
}\quad
{
 \includegraphics[angle=180,width=0.8\textwidth,trim= 0mm 0mm 0mm 0mm,clip]{Figs/on.ps}
 }
\caption{Two-minute (205 rotation) pulse stacks exhibiting PSR J0941--39's ``RRAT'' (off) and ``Pulsar'' (on) phases (\S\ref{sec:0941}, \ref{sec:explore}). The integrated profile shape is shown in the panel the the right. Tracks of the signal-to-noise ratio of the peak of each pulse with ${\rm SNR}>3$ are plotted below the pulse stacks. The brightest peaks of the on- and off-phases are comparable, indicating that the ``off'' state is not due to interstellar scintillation.}
\label{fig:0941}
\end{centering}
\end{figure*}

\subsubsection{PSR J1534--46}
This object appears to be intrinsically modulated (and nulling) with
single pulses appearing at 12-14 times the rotation period.
The modulation power spectrum shows a peak at $\sim4$ times the pulse period
near the centre of the integrated pulse. The single-pulse emission from this
pulsar is shown in Figure \ref{fig:stacks15}. The pulsar was detectable
in a Fourier search, giving $r=0.66$. We have not performed follow-up observations of
this source.

\subsubsection{PSR J1610--17}
Only one pulse was detected from this location, however there have been no
follow-up observations on the object. The original detection had a SNR of 7.4,
and no other detections above SNR $=4$ appeared in the 4.4 minute
observation. The single detected pulse is shown in Figure
\ref{fig:singlepulsers16}, and like PSR J0923--31, shows a frequency sweep
that agrees well with the expected frequency delay due to interstellar
dispersion, implying a astrophysical, galactic origin of the pulse.

\subsubsection{PSR J1647--36}
This pulsar appears to exhibit periodic nulling with bursts lasting on the
order of 1-6 pulse periods spaced by nulls lasting $\sim 27$ seconds (130
rotations). With each period of ``on'' activity, the bursts arrive at a phase
offset by \mbox{$5-10$\%} from the previous burst episode; it is possible that
the system is either strongly drifting or mode-changing, however follow-up
observations of this source are necessary to assess these hypotheses.

\subsubsection{PSR J1654--23}
The original detection showed 3 narrow (sub-0.5~ms), faint \mbox{(SNR $<7$)}
bursts which allowed the fit reported in Table \ref{table:newstuff}. The duty
cycle is the smallest observed to date for a pulsar, however due to the narrow
pulses and weak emission it is possible that some pulses were not visible
above the noise. It is possible that the period reported for this pulsar could
easily be an integer multiple of the true period, the effect of which would be
to broaden the duty cycle by the same integer value. Follow-up observations of
longer duration than the discovery pointing and at a faster sampling rate will
reveal more completely the nature of this pulsar.

\subsubsection{PSR J1753--38}
The period and DM for this object were derived by extracting and fitting just
the three bright pulses detected in the original pointing. Follow-up pointings
(totalling 21 minutes) revealed further pulses consistent with the flux,
period, and DM of the original detections.

\subsubsection{PSR J1753--12}
The discovery pointing containing this object was badly affected by
interference. This and the fact that only one very strong impulse was present
made genuine pulse extraction difficult, however three dispersed pulses from
the object were identified and parameters estimated. We have not yet made
follow-up observations of this source, however cleaner data may reveal more
frequent low-level emission.

\subsubsection{PSR J1825--33}
In the discovery pointing, this pulsar emitted seven bright, sequential pulses
near the beginning of the observation but was not visible thereafter, nor was
any emission visible in 25 minutes of follow-up observations. The profile
structure of this pulsar has a double peak (Figure \ref{fig:viewsus} shows one
pulse from this object).

\subsubsection{PSR J1850+15}
This pulsar's discovery pointing contained strong periodic interference which
probably prevented its detection in the original Fourier search ($r=1.04$;
interference is visible atop the fainter pulses of the object in
Fig. \ref{fig:stacks18}). The emission appears to be steady, with no evidence
for clear intermittent behaviour, although there is strong pulse-to-pulse
amplitude modulation. We have not performed follow-up observations of this
source.

\subsubsection{PSR J2033+00}
This source appears to be an intermittent pulsar with a high nulling fraction
and is sometimes detectable in the Fourier domain. There is some evidence of
sub-pulse drift; individual pulses rarely come from the same phase for more
than a few rotations (see Figure \ref{fig:stacks20}). The source was
detectable in single pulses for the entirety of the 25 minute follow-up on the
source, however became distinctly brighter (and weakly detectable in a Fourier
search; $r=1.86$) in the final 14 minutes.

\begin{figure*}
\begin{centering}
\begin{tabular}{m{1cm} m{18cm}}
\centering{(a) \mbox{Pulsar}}& \includegraphics[width=0.9\textwidth,trim= 45mm 0mm 0mm 0mm,clip]{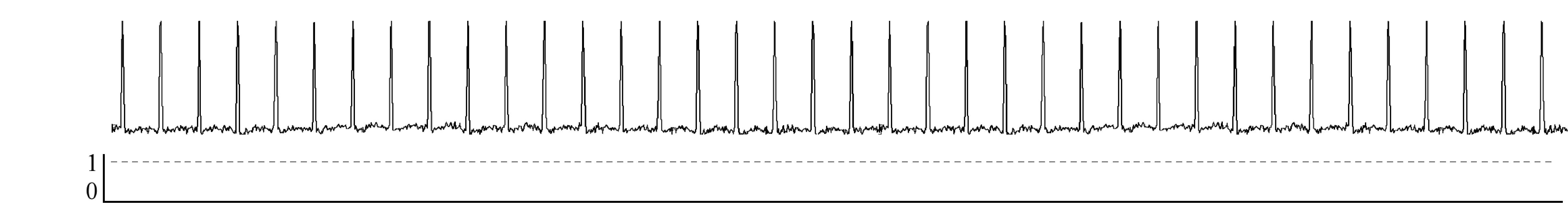}\\
\centering{(b) \mbox{Nulling} \mbox{Pulsar}} & \includegraphics[width=0.9\textwidth,trim= 45mm 0mm 0mm 0mm,clip]{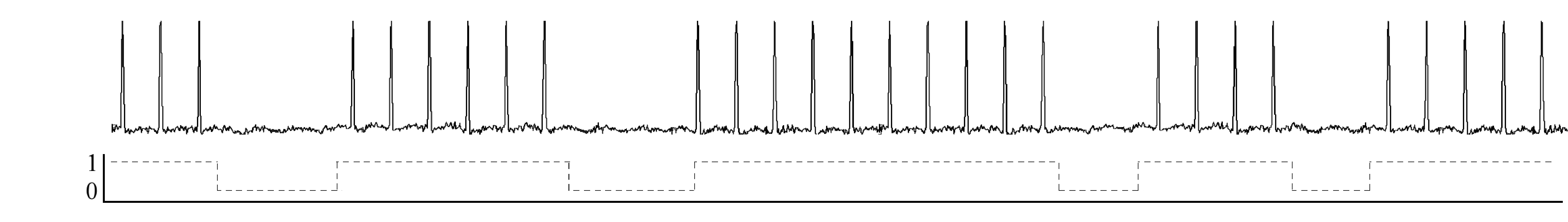}\\
\centering{(c) \mbox{Extreme} \mbox{Nulling} \mbox{Pulsar}} & \includegraphics[width=0.9\textwidth,trim= 45mm 0mm 0mm 0mm,clip]{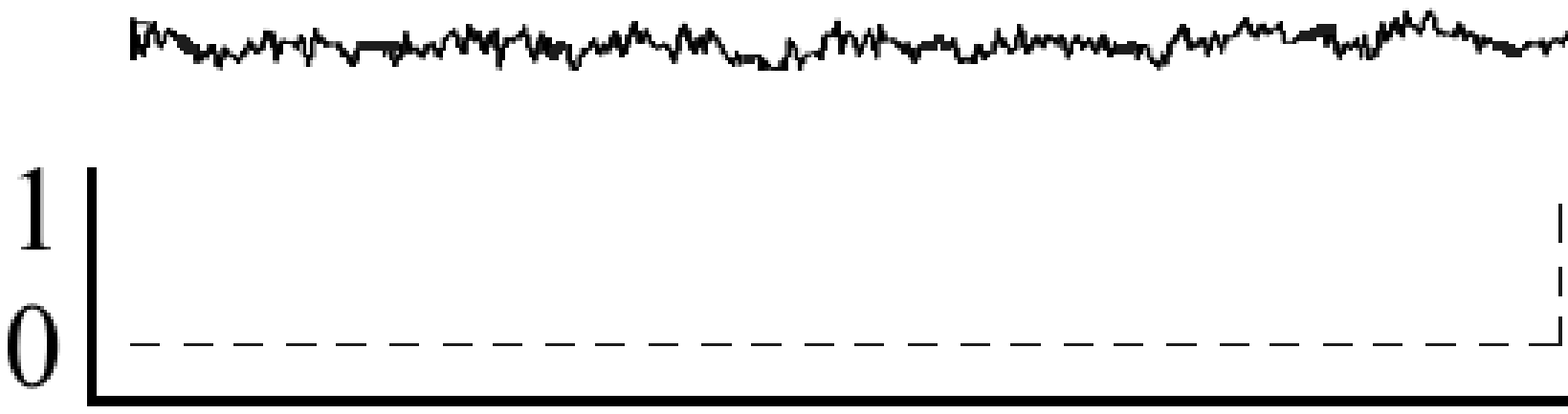}\\
\centering{(d) \mbox{RRAT}} & \includegraphics[width=0.9\textwidth,trim= 45mm 0mm 0mm 0mm,clip]{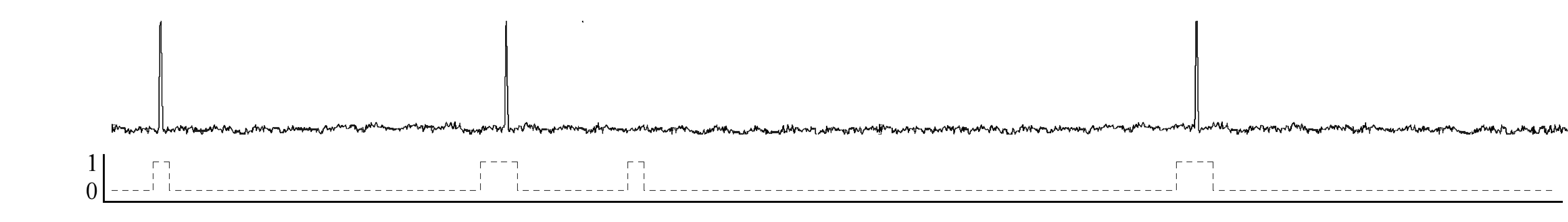}
\end{tabular}
\caption{Windowing as a descriptor for various emission observed in pulsars, where the average timescale of the window function defines the behaviour observed: (a) shows a regular pulsar, (b) is an example of a nulling pulsar, (c) an example of what we term an ``extreme nuller'' (e.g. nulling fraction $>75$\% of pulses are nulled and the {\it on}-window typically endures for a few times the rotation period), (d) an RRAT, in which the typical {\it on}-window has a duration of less than the pulsar period.}
\label{fig:windows}
\end{centering}
\end{figure*}

\section{discussion}\label{sec:discussion}
\subsection{Pulsar transient varieties and the nature of RRAT emission}

Regular pulsars are themselves intermittent radio sources, showing detectable
emission only for the fraction of time that their beams sweep Earth. The
individual pulses, however, are observed to show intrinsic pulse-to-pulse flux
and shape differences, the origin of which is yet unknown (though various
theories exist). The variations observed in pulsars can be broken down into
two main types, being: {\it pulse-to-pulse modulation}, in which the object
emits continuously, however at an intrinsically modulated amplitude, and {\it
nulling}, during which the pulsar beam is somehow quenched, emitting no
detectable radiation for a period of time. In many pulsars, nulling appears to
be a real phenomenon (with no underlying emission when in null states), with
nulls observed to occur from zero to 95\% of a pulsar's observed rotations
\citep{wang}. Some pulsars appear to null for days-weeks, with a period
derivative that is dependent upon the emission state \citep{sometimesapulsar}.

The range of nulling fractions observed in pulsars is easily described by a
windowing function, in which the number of sequential pulses observed at a
given time is defined by an on-window of some length. In this description,
regularly emitting pulsars have an indefinite on-window, while pulsars with
increasing percentages of nulling have on-windows of increasingly short
timescale, with extreme nullers (nulling fraction $\gtrsim75$\%) becoming more
likely to be detected in single pulse searches than periodicity
searches. Various degrees of pulsar windowing are illustrated in Figure
\ref{fig:windows}. It is clear that one could incorporate the RRATs into the
general pulsar population in this way, by just extending the non-nulling
window to a time of length $t<P$, causing only one pulse to be observed at a
time. It is a possibility that in this model, pulsars start off as
continuous emitters, and gradually increase their nulling fraction until they
become RRATs, which ultimately evolve further to emit no radio emission.

The alternative view of RRAT emission, related to pulse-to-pulse modulation,
was suggested by \citet{patrick06}, who indicated that if PSR B0656+14 were
observed at several times its distance from Earth, the highly variable single
pulses (emitting at up to 420 times the average pulse flux) could emulate the
single pulse behaviour observed by M06's distant discoveries. The
pulse-energy-distribution analysis of K09, however, makes an argument against
this hypothesis (although the analysis was not done on ``classic RRATs'' as we
define them below). It appears obvious that there will ultimately be some
fraction of RRATs that belong to this class, but deep integrations would be
expected to show underlying emission; the real issue, therefore, is in what
fraction of RRATs belong to this class, and if some do not, what are the
properties of those sources that do not?

Whether an effect of modulation or nulling, the current competing theories of
RRAT emission can be broken into three possibilities, in which the RRATs
represent:
\begin{itemize}
\item[(1)] a distinct class (intrinsic life-long single-pulsing) 
\item[(2)] faint pulsars with extreme pulse-to-pulse modulation
\item[(3)] an evolutionary phase of pulsars, e.g. where the magnetic field degrades or the spin period increases,
leading to increased nulling fractions
\end{itemize}
Scenarios (1) and (3) intertwine similarly with questions about pulsars observed to null (relating to whether nulling is an inherent phase of all pulsars or an effect of the local pulsar environment). The clarification of RRATs' nature not only has innate value, but the three scenarios above have quite different implications for how the RRAT birthrate integrates into that of pulsars, and how they contribute to the imbalance of core-collapse supernova rates and neutron star birthrates noted by \citet{evan}. For instance, in scenario (1) the latent RRAT population may actually exceed the pulsar population. Scenarios (2) and (3), on the other hand, are in part solutions to the birthrate problem, in that the birthrates contributed by RRATs are effectively absorbed into those of pulsars. If RRATs are simply distant, highly modulated pulsars, the sources detected as RRATs are simply abnormalities among the distant, otherwise undetectable pulsars sitting below the average luminosity limit of periodicity surveys; these distant and low-luminosity pulsars are accounted for in pulsar birthrate analyses \citep[e.g.][]{lorimerandfriends} with statistical estimates, using luminosity-dependent scaling factors. 

Although the original RRATs of M06 have a number of apparent properties which set them apart from regular pulsars, as detailed in Sec. \ref{sec:explore}, we define the {\it classic RRAT} as an object which emits only non-sequential single bursts with no otherwise detectable emission at the rotation period. When using the term {\it nulling pulsar}, we refer to an object which emits for some fraction of its rotations, however the emission coming in bursts of several sequential pulses. Finally, we refer to all objects which emit pulses for less than $\sim$25\% of their rotations as {\it extreme pulsar transients}.

Below we discuss the properties of our new transients and compare them with the RRATs of M06 and with the properties of the general pulsar population.

\subsection{A breakdown of our discoveries}\label{sec:varieties}
What is immediately striking about the 14 objects detected in this search is
the variety of dormancy timescales, covering a broad range of the nulling and
modulation phenomena related above. For some objects, such as PSR J1850+15,
the lack of previous detection by Fourier search techniques can be explained
by the presence in the data of intense periodic RFI, which lowers sensitivity
to fainter periodic sources. The pulsars PSR J0735--62 and PSR J1850+15 appear
to simply be regular pulsars which were likely missed in previous searches due
to the modulation and long period of PSR J0735--62, and the presence of
interference in the discovery pointing of PSR J1850+15.

The ``classic RRATs'' in our search include PSRs J1129--53, J1226--32,
J1654--23, J1753--38, and J1753--12. We also place the objects from which only
one pulse was observed in this category; although we have not detected further
events from the sources, the pulse widths are typical of pulsars and the
pulsation rates implied by our observations are consistent with the range of
pulsation rates observed in the RRATs published by M06, which ranged from one
per four minutes to one per three hours. We note that the bright single pulses
emitted by PSRs J1226--32 and J1753--38 are too bright and have a pulsation
rate too high to represent a long tail of a log-normal pulse energy
distribution, immediately indicating that scenario (2) above (extreme
modulation) is not the case for all RRATs.  This is less clear in the case of
PSR J1753--12, and we note that our sources may represent a mix of highly
modulated low-luminosity pulsars, and true RRATs.

Apart from PSR J0941--39, the remaining objects appear to be nulling pulsars with high nulling fractions. Only two of these objects were in an ``on'' state for a sufficient fraction of observing time to be Fourier-detectable. Additionally, it is difficult to say whether the source PSR J1825--33 is truly a nulling pulsar, or whether this source had an event which briefly activated its outer gap, as we have since not detected any further emission from the object.

PSR J0941--39 does not fit cleanly into any currently identified pulsar category, sometimes acting like a single-pulse-emitting RRAT, and sometimes like a nulling (though not extreme-nulling!) pulsar. This object may play an important role in understanding RRAT emission.

\subsection{Exploration of RRATs and the properties of our new discoveries}\label{sec:explore}
This search represents an important step in understanding extreme pulsar
intermittents in several ways. First, we have surveyed the largest area to
date for such sources in a region off the galactic plane, and added the
discovery of 12 highly intermittent pulsars to a small known population of
such objects. Given that neutron stars are mainly born from massive stars in
the plane of the galaxy, the high galactic latitudes predispose us to sample a
population which has had a sufficiently high kick velocity and sufficient time
to reach its current position. Also, the automatic RFI rejection and
additional candidate visualisation techniques appear to have given us a higher
sensitivity to low DM single pulses, as seen by the fact that M06 discovered
no sources at \mbox{DM $<88$~pc\,cm$^{-3}$} while we have discovered sources
at a broad range of DM, with twelve of the fourteen detections at less than
\mbox{88\,pc\,cm$^{-3}$}, and with previously known pulsars positively
identified down to \mbox{${\rm DM}=3$\,pc\,cm$^{-3}$.} Additionally, the latitude
ranges of past RRAT searches (K06, D09, K09) did not exceed 5\,degrees. This
increased DM sensitivity and high, complementary latitude range makes our
discoveries salient objects for assessing the inherent properties of nearby,
RRAT-like bursting sources and how neutron stars with extreme intermittence
relate to the Galactic population of pulsars.

Throughout this discussion we analyse all of our sources which are
sufficiently intermittent to not be detectable in a Fourier search---that is,
the ten sources not plotted in Figure \ref{fig:stacks}. Where appropriate, we
distinguish the analysis of ``classic RRATs'' and that of ``nulling pulsars.''
For reference we also include the M06 RRATs and the four D09 objects which, as
noted by their study, did not show detectable underlying regular
emission. Additionally, we include the five objects which were not noted by
K09 to exhibit Fourier-detectable emission or clear nulling behaviour.

The sparse emission and dissimilarity of magnetic field strength to pulsars
which emit giant pulses on $\mu$s timescales \citep[][]{cognard} are what most
definitively led McLaughlin et al. to the conclusion that their sources were a
new neutron star population, however several other of the observed properties
of the original 11 sources also distinguish them from the general pulsar
population. This includes their low duty cycles, high single-pulse
luminosities, and their tendency to be distant ($>2$~kpc, which may have
partially been due to a selection effect; this particular RRAT feature gives
more weight to the claims of \citealt{patrick06} that RRATs may simply be
distant, highly modulated pulsars). To further our understanding of the
properties of RRATs, here we explore their various distinguishing features in
comparison to the extreme intermittents detected in our search.

The galactic distribution of our new objects (of the classic RRATs and of our 14 discoveries in general) agree well with the distribution of pulsars, concentrated more highly towards the Galactic plane and the Galactic centre. Though not statistically rigourous with a sample of fourteen objects, the galactic $z$-height distribution ($z=d\cdot{\rm sin}(l)$) of our discoveries above $z=200$~pc is well-fit with an exponential and a scale height of 300~pc, in good agreement with the galactic pulsar distribution. This agreement and the discovery of RRATs in the high latitude surveys indicate that the RRATs are not a young phenomenon confined to the galactic plane like anomalous X-ray pulsars \citep[e.g.][]{deadant}. The object of second largest $z$-height in our sample is the RRAT J1610--17 with a scale height of $z=0.86$~kpc. With a birth at $z=0$~pc and a proper motion outward from the galactic plane at 100~km\,s$^{-1}$, the implied age of this object is $\gtrsim10^7$ years.

\begin{figure}
{
 \includegraphics[angle=270,trim=5mm 15mm 0mm 1mm,clip,width=0.475\textwidth]{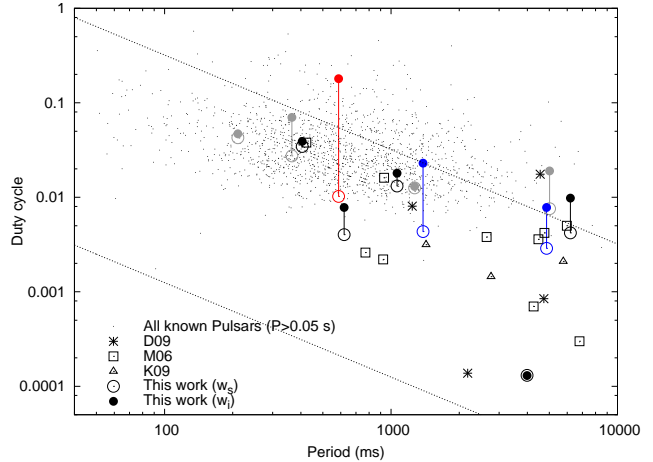}
}

\caption{The duty cycle versus period for various detections (as detailed in \S\ref{sec:varieties}). Among the sources presented in this paper, pulsars are highlighted in blue, RRATs in black, nulling pulsars in grey, and PSR J0941--39 is represented by the red symbols. The duty cycles of our detections are shown in two ways: using the integrated pulse width (measured after all available on-pulses were stacked at the pulsar period) and using the width of the brightest single pulse. M06, D09, and K09 reported only single-pulse widths. The dotted lines represent traces of $w=0.125$~ms and 32~ms, beyond which our search sensitivity to individual pulses drops significantly.}
\label{fig:dutycycles}
\end{figure}

Figure \ref{fig:dutycycles} plots the duty cycles of known pulsars and the objects analysed here. As expected, the single pulse widths, $w_{\rm s}$, gleaned from our new objects lie primarily within the optimal sensitivity range set by the box-car matched filters in our search, as described in section \ref{sec:methods}.

The most notable feature of this diagram is that upon plotting our integrated pulse width measurements (filled circles) with the single pulse width measurements (open circles), it becomes apparent that when using duty cycles measured from individual pulses (as were used in the calculations of M06, D09, and K09), the values measured for our RRAT-like objects become systematically lower by a factor of $\sim$3, and up to a factor of more than 10 for the extreme case of J0941--39. There may be several causes of this; for pulsars with nulling, pulse drifts and mode changes are often observed, which can work to broaden the integrated pulsar profile (illustrated well by PSR J2033+00's pulses in Fig. \ref{fig:stacks20} above). Multi-component integrated profiles, made up of a superposition of otherwise narrow pulses emitted from a range in latitudes, can also broaden an object's duty cycle between single and integrated pulse measurements (as has occurred in the extreme case of PSR J0941--39 with its triple-peaked profile).
It is clear that were an integrated pulse width measured for the sources of M06, D09, and K09, the duty cycle distribution of RRATs would more closely resemble that of the general pulsar population, instead of appearing to sit characteristically lower than pulsars (the composite profiles for two RRATs shown in \citet{mauranewpaper} verify this suggestion). As the RRAT and pulsar duty cycle distributions are similar, it is probable that the beaming fraction of such objects will likewise be akin to that of pulsars.


\begin{figure}
\begin{centering}
 \includegraphics[angle=270,trim=3mm 7mm 0mm 7mm,clip,width=0.45\textwidth]{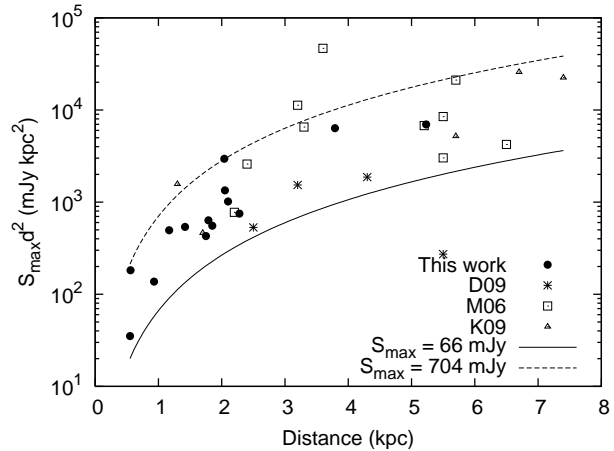}
\caption{Dependence of luminosity on the distance of the sources. The lower curve represents our single pulse flux sensitivity limit for a pulse width of 32~ms, while the upper curve shows a representative limit above which a highly modulated pulsar would become detectable in a Fourier search, as detailed in \S\ref{sec:explore}}
\label{fig:lvd}
\end{centering}
\end{figure}

Next we examine the distribution of the single-pulse luminosities, as shown in
Figure \ref{fig:lvd}. The points cover a broad range in both distance and
luminosity, appearing to show a downturn with proximity to the observer. The
lower single-pulse sensitivity limit of our search obeying $L\propto d^2$ is
plotted (at 66~mJy for our 32~ms filter). In light of the possibility that
scenario (2) (modulation) is true, we consider the following: Fourier
transform searches for SNR~$\geq 8$ pulses as done by \citet{ED} over the 4.4
minute survey pointings have a sensitivity limit to pulsars (given a typical
duty cycle of 5\%) with an average single-pulse peak flux of $\langle S_{\rm
peak}\rangle \geq 3.3$~mJy. The prominence and statistics of extremely
modulated pulsars (like PSR B0656+14) are not yet well characterised, however
the study of \citet{patrick06a} of 187 pulsars at 1~GHz revealed that typical
modulation rarely exceeds ten times the average pulse energy (with pulses
exceeding this range typically lasting nanoseconds, and called ``giant
pulses'', \citealt[e.g.][]{cairns04}). Their follow-up study of PSR B0656+14,
as previously noted, revealed infrequent single pulse emission peaking up to
420 times the average single pulse intensity. Here we consider that such
extreme sources will be relatively uncommon, and take a population of
hypothetical sources bursting at half the maximum $S_{\rm max}/\langle S_{\rm
peak}\rangle$ ratio of B0656+14, that is, a bursting population with typical
$S_{\rm max} = 210\langle S_{\rm peak}\rangle$. If such a population is
contained in our data, sources above the limit of $\langle S_{\rm peak}\rangle
\geq 3.35$~mJy will have bursts at $S_{\rm max} \simeq 700$~mJy. Sources
brighter than this limit will be detectable in Fourier searches and be tagged
as pulsars, however below this limit the periodic emission will have been
missed, with only the brightest bursts detectable in single pulse
searches. This limit is plotted in Fig. \ref{fig:lvd}, and interestingly,
appears to correspond to the upper envelope of the maximum source luminosities
found by our single pulse search.

The natural interpretation of this is that the nearby, highest-luminosity sources exhibiting erratic emission had bright enough low-level emission that instead of being detected as nulling pulsars or RRATs, they were discovered in periodicity searches and taken into the population as pulsars (albeit modulated or nulling ones). 
When a similar test is done for the RRATs found by M06 (whose longer, 35.5 minute pointings have higher sensitivity to periodic signals and a higher likelihood of containing the rarer brightest outbursts of a highly modulated pulsar; a periodicity search of that survey is described in \citealt{pksmb}), two sources, PSRs J1317--5759 and J1819--1458, still lie significantly above the limit of the Fourier analysis. In the framework of our interpretation, these two sources are therefore either examples of extreme sources among modulated pulsars (however nearby equivalents among the known pulsar population, apart from B0656+14, have yet to be recognised), or alternatively they are simply examples of genuine RRATs, where the intrinsic pulsar emission is quite strong, however usually invisible because of a very high nulling fraction.

\begin{figure}
\begin{centering}

 \includegraphics[angle=270,trim=5mm 4mm 0mm 5mm,clip,width=0.45\textwidth]{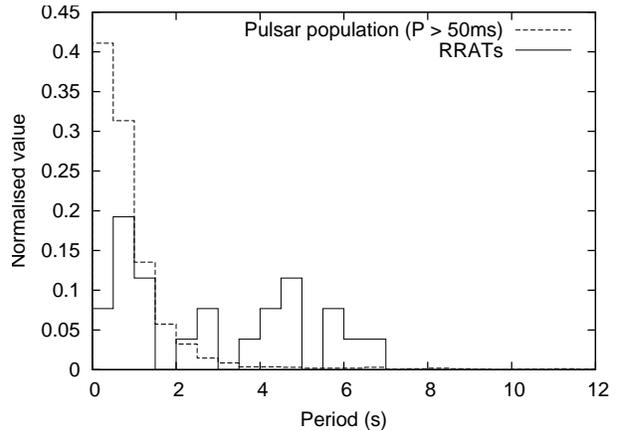}
\caption{Normalised period distributions for pulsars and the single-burst emitting RRATs of M06, D09, K09, and this survey.}
\label{fig:periods}
\end{centering}
\end{figure}

Among the four of our RRATs for which we could measure periods, although very
statistically poor, the previously noted RRAT feature---that they have a
tendency to longer periods than pulsars---is observed in our new objects,
where 50\% have periods longer than one second. Plotted in Figure
\ref{fig:periods} are the normalised distributions for our and the published
``classic RRAT'' sources as compared with the known pulsar population for
pulsars with $P>50$~ms (to restrict the analysis to non-recycled
pulsars). When a Komolgorov-Smirnov test is run on the un-normalised
distributions, it results in a high probability, $P=0.999985$ ($>4\sigma$),
that the distributions are not drawn from the same parent population. Few
pulsars have periods greater than $P = 3$~s and because sources such as
anomalous X-ray pulsars/magnetars/soft gamma-ray repeaters all have periods
greater than a few seconds, it is tempting to draw an association between such
sources and RRATs; the recent release of timing data on six RRATs by
\citet{mauranewpaper} show a further similarity of RRATs to these other
neutron star types in their placement in the period-period derivative diagram,
and a similar result in periodicity distribution differences. However, the
broad distribution of source periods and the presence of 7 RRATs with periods
below one second suggests that a more basic factor in the population, for
instance age, which can create the observed tendency to longer periods. For
the RRATs which have had timing analysis done, all appear to have ages of a
few times $10^{5-6}$ years. However, the statistical analysis of
\citet{mauranewpaper}, which compared various derived RRAT and pulsar
quantities, does not indicate a strong difference in the distribution of RRAT
and pulsar ages, though does indicate a strong probability that the
distribution of RRAT and pulsar magnetic field strengths are significantly
different. There appears to be no relation between RRAT period and distance
from the galactic plane (a proxy, albeit a loose one, for pulsar age).

The source PSR J0941--39 contributes a peculiar voice to this discussion,
appearing for hours, to possibly weeks at a time like a classic
RRAT---erratically emitting single, non-sequential pulses at a relatively
constant rate---while at other times appearing as an intense ($\sim 10$ mJy)
pulsar with a $<10$\% nulling fraction, extraordinary sub-pulse drift, and
latitude-dependent modulation. The object appears superficially to tie a close
link between nulling phenomena and RRATs, possibly signifying a pulsar which
is in some way transitioning from a pulsar mode to an RRAT mode. If this is
the case, the source provides a link both between RRATs and pulsars as well as
a link to older, non-radio-emitting neutron stars---i.e. that the RRAT phase
represents a degradation of the radio emission mechanism before advancing
beyond the radio pulsar death line. As a timing solution is acquired for this
pulsar, it will be interesting to note whether it is of an advanced age. It is
rather surprising how luminous this pulsar is, and what a small nulling
fraction it possesses when in its ``pulsar'' state. It is important to note
that the site of the RRAT emission is coincident with the three peaks of the
profile when the pulsar is on. Thus, the RRAT emission is from the same site
as the normal pulsar emission, again implying similar beaming fractions for
the two states.


The properties of the new transients found in this search, and particularly of the five new RRATs, give various indicators that RRAT behaviour is well described by either or both of scenarios (2) or (3) above. 
The idea that observed RRAT behaviour may be an observed consequence of extreme ($>99$\%) nulling is supported by the high nulling fraction of the objects tagged as ``nulling pulsars'' above; while previously the ``record'' nulling fraction was held by PSR B1713--40 at 95\% \citep{wang}, PSRs J1825--33, J1647--36, and J1534--46 all have implied nulling fractions of more than $97$\%, telling us that these objects may imply a  ``bridging population'' of high-nulling-fraction pulsars, while objects with higher nulling fractions (smaller and less frequent on-windows) exhibit RRAT behaviour. A full statistical nulling fraction analysis of the \citet{ED} and \citet{BJ} pulsars may indicate that the nulling fraction of the RRATs discovered in this search fall naturally in the tail of pulsars with significant nulling fractions shown to exist by \citet{wang} in other surveys.
This proposition, along with the lack of RRAT features that are physically distinct from pulsars (except their period distribution)---including their galactic and duty cycle distribution---point to a strong relationship to the ``normal'' radio pulsar population. 
The story is complicated and strengthened by the apparent mode-switching of PSR J0941--39, which lends significant support for scenario (3) in that it is clear that the object in its RRAT mode is not simply a modulated pulsar, and it is likewise clear that the object is both RRAT and pulsar in nature, signifying a transition between the two populations and more speculatively, implying that nulling pulsars may undergo stepped increases in their nulling fraction with age. The analyses of \citet{wang} indicate no correlation between nulling fraction and characteristic pulsar age, however the number of pulsars exhibiting nulls appears to increase with proximity to the radio death line (see their Fig. 8). If all RRATs are relics of nulling pulsars, the broadening of integrated pulse measurements supports the fact that despite their sparse emission, the emitted pulses may still be exhibiting drifting or mode-changing phenomena.


\section{Summary}\label{sec:summary}
We have re-examined the intermediate galactic latitude surveys of \citet{ED} and \citet{BJ} for single pulses of radio emission of duration $0.125<w<32$~milliseconds. This search revealed 14 intermittent sources which appear to be neutron stars. Periods ranging 0.2 to 6.2~s were measured for 12 of these sources. While only four of the objects could be detected retrospectively by Fourier search techniques, the remaining 10 exhibited intermittent emission behaviour akin to both extreme nulling pulsars and the classic RRATs discovered by \citet{mclaughlinetal06}. The sensitivity of our search techniques to bursts of lower dispersion measure allowed us to find objects which were closer than those detected by previous searches.

The general properties of the new sources appear to paint details of the extreme intermittent and RRAT population that concur with having similar emission modes and galactic distribution to pulsars, despite being significantly less emissive than the general population. It was found that the duty cycle distribution of our detections mimics that of pulsars. These similarities, and the discovery in this search of pulsars with extremely high nulling fractions, lead us to suggest that RRATs may represent an extreme type of nulling pulsar whose window of on-activity is shorter than the rotation period of the object, and that the sources observed to exhibit RRAT behaviour are a natural extension of the tail of pulsars observed with high nulling fractions. Furthermore, PSR J0941--39 appears to oscillate between a clear ``nulling pulsar'' state and a more intermittent ``RRAT'' mode. This source may represent the first detection of an object in transition from a bright pulsar with an average nulling fraction to a single-bursting source, closely tying nulling pulsars to RRATs and indicating a possible evolutionary progression. Further studies and timing are warranted and upcoming on this object.

We conclude by noting that identifying RRATs is not a simple task, as they may represent a mixed population of intrinsically single-pulsing pulsars and highly modulated pulsars whose brightest single pulses singly exceed the system noise.
Modulated pulsars appearing as RRATs are expected to be few, as among the local population of pulsars, PSR B0656+14 is currently the only pulsar known to be as modulated as the population required for RRAT emission.
Nearby pulsars which are part of such a population may in the future be identified by pulse energy distribution studies of nearby highly modulated pulsars, using long observations which have a chance of detecting the rarer brightest outbursts from such objects. Statistical studies of these pulsars may help separate ``true'' RRATs from simply highly modulated pulsars.

To discern the nature of RRATs, other extreme pulsar intermittents, and highly modulated objects, it will be critical in coming years to both identify more such sources, and to continue timing analyses on the known sources to determine their relationship to other neutron star populations. Of additional interest are the statistics of intermittent phenomena in extreme nulling/modulated pulsars among the known pulsar population, the analysis of which may reveal further insight into the relationship between regular pulsars and the vast variety of intermittent behaviours observed in these equivocal neutron stars.

\section{Acknowledgements}
We would like to thank R. Bhat and W. van Straten for useful discussions about pulsar phenomena and software during this project and the preparation of the manuscript.

\end{document}